\documentclass[prb,twocolumn,showpacs,showkeys,secnumarabic,amssymb,amsmath,unsortedaddress]{revtex4-1}
\usepackage{amssymb,amsmath}
\usepackage{lipsum}
\usepackage{mathtools}
\usepackage{color}
\usepackage{graphics}
\usepackage{graphicx}
\usepackage{tabularx} 
\usepackage{enumitem}
\usepackage[dvipsnames]{xcolor}
\usepackage{bm} 
\usepackage{soul}
\graphicspath{{pics/}}
\begin{document}

\def \rg{\rangle}
\def \vac{|\text{vac}\rangle}
\def \vacc{\langle\text{vac}|}

\title{Time dynamics of Bethe ansatz solvable models}

\author{Igor Ermakov}
\affiliation{Skolkovo Institute of Science and Technology, Ulitsa Nobelya, 3, Moskva, Moscow Oblast, 143026}
\affiliation{Steklov Mathematical Institute of Russian Academy of Sciences, Gubkina str. 8, Moscow 119991, Russia}
\affiliation{New York University Shanghai, 1555 Century Avenue, Pudong, Shanghai 200122, China}

\author{Tim Byrnes}
\affiliation{New York University Shanghai, 1555 Century Avenue, Pudong, Shanghai 200122, China}
\affiliation{Department of Physics, New York University, New York, NY 100003, USA}

\begin{abstract}
We develop a method for finding the time evolution of exactly solvable models by Bethe ansatz.  The dynamical Bethe wavefunction takes the same form as the stationary Bethe wavefunction except for time varying Bethe parameters and a complex phase prefactor.  From this, we derive a set of first order nonlinear coupled differential equations for the Bethe parameters, called the dynamical Bethe equations.  We find that this gives the exact solution to particular types of exactly solvable models, including the Bose-Hubbard dimer and Tavis-Cummings model. These models go beyond the Gaudin class, and offers an interesting possibility for performing time evolution in exactly solvable models. 
\end{abstract}


\maketitle

%
%
%
\section{Introduction}
\label{Intro}

Exact methods of mathematical physics have substantially pushed our understanding of many paramount nonlinear phenomena. One such method is the Quantum Inverse Method (QIM) which was developed almost 40 years ago by Faddeev, Sklyanin, Takhtadzhyan and others \cite{sklyanin1979quantum,sklyanin1982quantum,kulish1982quantum}. QIM together with the algebraic version of Bethe Ansatz \cite{slavnov2018ABAPre,levkovich2016bethe} has been successfully applied to various problems from different areas of physics such as one dimensional BECs \cite{lieb1963exact,knap2014quantum}, spin chains \cite{maillet2007heisenberg,kitanine1999form,kato2003next,bortz2005exact}, $(1+1)$ models of quantum field theory \cite{faddeev1982integrable}, $(2+1)$ model of classical statistical physics \cite{thiery2016exact}, conformal field theory and string theory \cite{arutyunov2004bethe}, quantum optics \cite{bogoliubov2012exactly}, and quantum dots \cite{bortz2007exact}.


Obtaining the time dynamics of quantum many-body systems remains an important but very challenging problem due to the high computational and calculational demands.  In the case of the QIM, the dynamics of the system after a quench of one or several parameters has been successfully shown \cite{faribault2009bethe,zill2018quantum}. However, in general, QIM without modifications can not be applied to the system with time-dependent parameters. Recently several exact methods for time-dependent Hamiltonians were proposed. In Ref. \cite{sinitsyn2018integrable}, a set of conditions under which the Schrodinger equation can be solved exactly was presented. It was also shown in Ref. \cite{sinitsyn2018integrable} that among Hamiltonians satisfying these conditions are the multistate Landau-Zener model and the generalized Tavis-Cummings model. Earlier in Ref. \cite{barmettler2013non}, Barmettler, Fioretto and Gritsev proposed a generalization of the Bethe wavefunction for the dynamical case and presented its explicit form for the detuning driven Tavis-Cummings model. In Ref. \cite{fioretto2014exact}, by means of correspondence between the class of Gaudin models and the classical Knizhnik-Zamolodchikov equations some exact solutions for Gaudin-magnets were obtained for special choices of time dependence of the coupling constants. There has also been progress in studies of the exact dynamics of periodically driven systems \cite{gritsev2017integrable}. However, to our knowledge a general formulation of how to perform the time evolution of an integrable system has not been shown.  

In this paper, we study the generalization of Bethe wavefunction for the time-dependent case.  Specifically, consider that we are dealing with an integrable system with Bethe wavefunction
\begin{align}
\prod^M_{j=1}\mathbf{B}(\lambda_j)\vac, 
\label{generalB1}
\end{align}
where $\mathbf{B}(\lambda)$ is an operator which depends on complex parameter $\lambda$ and $\vac$ is the pseudo-vacuum  reference state, specific to the model being considered.  For an initial state that can be represented by the Bethe wavefunction, we show that its time evolution can be described using the dynamical Bethe wavefunction, 
\begin{align}
e^{ip(t)}\prod^M_{j=1}\mathbf{B}(\lambda_j(t))\vac,\label{generalB2}
\end{align}
where $p(t)$ is a complex phase. The time dependent wavefunction has exactly the same structure as Bethe wavefunction, but its parameters are functions of time and it has time varying prefactor. 
One of the most important features of the Bethe vectors is that it allows for the determinant representation for observables \cite{slavnov1989calculation}, which is widely used in calculations of the Bethe ansatz \cite{gamayun2018impact,bulchandani2018bethe}. The fact that the time-dependent wavefunction (\ref{generalB2}) has the structure of a Bethe vector allows us to transfer all the Bethe ansatz machinery to the time-dependent case.

When a system is exactly solvable by QIM, one can always make (\ref{generalB1}) an eigenfunction, by choosing special values of the parameters $\lambda_j$, which satisfy the Bethe equations. In this paper, we formulate a set of conditions for when the dynamical Bethe wavefunction (\ref{generalB2}) satisfies the time-dependent Schrodinger equation. The set of conditions is a set of nonlinear coupled differential equations, which we call the dynamical Bethe equations. The time-dependent wavefunction can always be represented by the dynamical Bethe wavefunction (\ref{generalB2}) for an arbitrary smooth time-dependence of the model parameters if the Hilbert space of the model under consideration is small enough. We provide an explicit example of the dynamical Bethe equations for a detuning driven Bose-Hubbard dimer. 

The form of the wavefunction (\ref{generalB2}) first appeared in Ref. \cite{barmettler2013non} for the Tavis-Cummings model, where the set of dynamical Bethe equations for $\lambda_j(t)$ was found, and its connection of trajectories $\lambda_j(t)$ with classical motion in a potential was established. So far all the examples of the dynamically integrable models considered in \cite{sinitsyn2018integrable,barmettler2013non,fioretto2014exact} belong to the Gaudin class \cite{gaudin1983fonction} of integrable models or models with a classical R-matrix. Here we show that  (\ref{generalB2}) can be applied to a wider class of integrable models, which goes beyond Gaudin class. The Bose-Hubbard dimer example that we show here belongs to the so-called rational XXX R-matrix class. We note that the set of conditions formulated in Ref. \cite{sinitsyn2018integrable} does not require that the model should belong to the Gaudin class to be dynamically integrable. Furthermore, models which can be solved by dynamical Bethe wavefunction also do not necessarily satisfy the set of conditions in Ref. \cite{sinitsyn2018integrable}.

The paper is organized as follows. In Sec. \ref{qim} we discuss the general procedure of constructing the dynamical Bethe wavefunction. In Sec. \ref{BHDimer} we derive the dynamical Bethe equations for a Bose-Hubbard dimer with driven detuning and quenching. In Sec. \ref{outlook} we summarize and discuss the future prospects of our method. For more background and details about the Bethe Ansatz technique and our derivations, we refer the reader to the Appendix.

\section{Dynamical Bethe equations}
\label{qim}

In this section we discuss the general method of finding the dynamical Bethe wavefunction, without specifying the model. For the reader who is not familiar with Bethe ansatz, we refer them to the Supplemetary Material, or for an extensive review see for example Ref. \cite{SlavnovLecturesArXiv}.

We first assume that the model under consideration can be solved by algebraic Bethe ansatz.  We also assume that the set of these Bethe vectors form a complete orthogonal set. This condition should be checked for every specific model separately, but for the vast majority of physically relevant models it is known to be satisfied. Also for simplicity we restrict the considered models to be those with a rational R-matrix and XXX or XXZ-like R-matrices. In practice these three classes cover most physically relevant models. 


A central quantity in integrable models is the trace of the monodromy matrix $\tau(\lambda)$ (see Appendix). This operator  has many useful algebraic properties provided by the integrability of the model, its specific form should be defined for each model separately. By construction $\tau(\lambda)$ is explicitly connected with the Hamiltonian $\hat{\mathcal{H}}$ of the model under consideration. Usually the Hamiltonian $\hat{\mathcal{H}}$ can be expressed as some elementary function or a residue of $\tau(\lambda)$ at some certain point $\lambda_0$. Because of the connection between $\hat{\mathcal{H}}$ and $\tau(\lambda)$, we will see that it is beneficial to consider the following Schrodinger-like equation
\begin{equation}
\label{Schrodinger_tau}
i\frac{d}{dt}|\Psi(t)\rangle=\tau(\lambda)|\Psi(t)\rangle .
\end{equation}
This will allow us to learn the complete information about the time dynamics of the system. 

We look for the solution of (\ref{Schrodinger_tau}) of the form
\begin{equation}\label{bWF_gen}
|\Psi^\sigma_M(t)\rangle=e^{ip^\sigma(t)}\prod^M_{j=1}\mathbf{B}(\lambda^\sigma_j(t))\vac,
\end{equation}
here $M$ is the number of excitations in the system and $\sigma$ enumerates the eigenstates. 
At $t=0$, the vectors (\ref{bWF_gen}) are eigenvectors which form a complete orthogonal set and the set of parameters $\Lambda^\sigma_M(t)=\{\lambda^\sigma_1(t),\lambda^\sigma_2(t),...,\lambda^\sigma_M(t)\}$ satisfies the stationary Bethe equations for each $\sigma$. We demand time-dependent wavefunctions to also form a complete set
\begin{equation}\label{Orth_dynam}
\sum_\sigma |\Psi(\{\lambda^\sigma(t)\})\rangle\langle\Psi(\{\lambda^\sigma(t)\})| \propto \hat{I} ,
\end{equation} 
where we have a proportionality because the wavefunctions are not normalized. The expansion of Bethe vectors (\ref{bWF_gen}) over a convenient basis is a difficult problem and in general not solvable, because of the complex structure of (\ref{bWF_gen}).  For example, $\mathbf{B}(\lambda)$ can be represented as a series of exponential length. 

Thus instead of studying the Schrodinger equation (\ref{Schrodinger_tau}) directly, we demand that
\begin{align}
\langle\Psi(\{\lambda^{\sigma'}(t)\})|\Psi(\{\lambda^\sigma(t)\})\rangle=0 ,\label{determ1}
\end{align}
for $\sigma<\sigma'$, where $\sigma\in [1,L]$ and $L$ is the dimensionality of the Hilbert space under consideration.
This states that the Bethe vectors are mutually orthogonal for all $ t $.  We also demand that 
\begin{align}
\langle\dot{\Psi}(\{\lambda^{\sigma'}(t)\})|\Psi(\{\lambda^\sigma(t)\})\rangle+\langle\Psi(\{\lambda^{\sigma'}(t)\})|\dot{\Psi}(\{\lambda^\sigma(t)\})\rangle=0 ,
\label{determ2}
\end{align}
which must be satisfied for any solution of (\ref{Schrodinger_tau}). 

We now would like to write (\ref{determ1}) and (\ref{determ2}) as a set of coupled differential equations.  
Eq. (\ref{determ1}) may re-expressed in this form by writing $|\Psi(\{\lambda^\sigma(t)\})\rangle$ in terms of its derivative, which for Bethe vectors always take a special form. 
To show this, we start by finding the result of operatoring $\tau(\mu)$ on the Bethe wavefunction (\ref{bWF_gen}), giving the 
well-known result
\begin{align}
\tau(\mu) & \prod^M_{j=1}\mathbf{B}(\lambda^\sigma_j)\vac=\Theta(\mu, \{\lambda_j^\sigma\}) \prod^M_{j=1}\mathbf{B}(\lambda_j^\sigma)\vac  \nonumber \\
& + \sum^M_{n=1}\phi_n(\mu, \{\lambda_j^\sigma\})\mathbf{B}(\mu) \prod^M_{j=1 \atop j\neq n}\mathbf{B}(\lambda_j^\sigma)\vac ,
\label{tau_acts_on_WF_art}
\end{align}
where $\Theta(\mu, \{\lambda_j^\sigma\})$ and $\phi_n(\mu, \{\lambda_j^\sigma\})$ are eigenvalues and the off-shell functions defined in (\ref{thetaL}) and (\ref{offshellfunc}) correspondingly. By combining (\ref{Schrodinger_tau}), (\ref{bWF_gen}),  and (\ref{tau_acts_on_WF_art}), we obtain
\begin{align}
\label{substitution_to_SH}
	& i(i \frac{dp}{dt}-\Theta(\mu, \{\lambda^\sigma_j(t)\})\prod^M_{j=1}\mathbf{B}(\lambda^\sigma_j(t))\vac \nonumber\\
	& = -i\frac{d}{dt}\prod^M_{j=1}\mathbf{B}(\lambda^\sigma_j(t))\vac \nonumber \\
	& +\sum^M_{n=1}\phi_n(\mu, \{\lambda^\sigma_j(t)\})\mathbf{B}(\mu) \prod^M_{j=1 \atop j\neq n}\mathbf{B}(\lambda^\sigma_j(t))\vac .
\end{align}
%
Demanding that the right hand side is proportional to the left hand side,  
	\begin{align}
	\label{proportionBW}
	& 	f(\{\lambda^\sigma_j(t)\})\prod^M_{j=1}\mathbf{B}(\lambda^\sigma_j(t))\vac = -i\frac{d}{dt}\prod^M_{j=1}\mathbf{B}(\lambda^\sigma_j(t))\vac \nonumber \\
 & 	+\sum^M_{n=1}\phi_n(\mu, \{\lambda^\sigma_j(t)\})\mathbf{B}(\mu) \prod^M_{j=1 \atop j\neq n}\mathbf{B}(\lambda^\sigma_j(t))\vac ,
	\end{align}
	%
%
where $f(\{\lambda^\sigma_j(t)\})$ is a smooth function. If (\ref{proportionBW}) is satisfied we can solve (\ref{Schrodinger_tau}) with (\ref{bWF_gen}) by choosing special form of phase factor $p^\sigma(t)$
\begin{equation}
\label{phase_factor}
p^\sigma(t)=-\int^t_0\left[i\Theta(\mu, \{\lambda^\sigma_j(t')\})+f(\{\lambda^\sigma_j(t')\})\right]dt'.
\end{equation} 
Although it is possible to explicitly find both $f(\{\lambda^\sigma_j(t)\})$ and $p^\sigma(t)$, in practice this is not necessary, because the phase factor $e^{ip^\sigma(t)}$ cancels for any observable due to normalization.

After substitution of (\ref{proportionBW}) into (\ref{determ1}), the conditions (\ref{determ1}) transfers to the set of differential equations:
\begin{align}
\label{determ1full}
& i\langle\Psi(\{\lambda^{\sigma'}(t)\})|\dot{\Psi}(\{\lambda^\sigma(t)\})\rangle = \nonumber \\
&  	+\langle\Psi(\{\lambda^{\sigma'}(t)\})|\sum^M_{n=1}\phi_n(\mu, \{\lambda^\sigma_j(t)\})\mathbf{B}(\mu) \prod^M_{j=1 \atop j\neq n}\mathbf{B}(\lambda^\sigma_j(t))\vac ,
\end{align}
Now conditions (\ref{determ2}) and (\ref{determ1full}) are set of $L^2-L$ nonlinear differential equations, with $ML$ variables, where $M$ is the number of parameters which parameterize Bethe wavefunction (\ref{bWF_gen}).  The solution of (\ref{determ2}) and (\ref{determ1full}) is a set of trajectories $\Lambda^\sigma(t)=\{\lambda^\sigma_1(t),...,\lambda^\sigma_M(t)\}$, for each wavefunction enumerated by $\sigma$. So when $M=L-1$ the number of equations coincides with the number of variables and (\ref{determ1}) and (\ref{determ2}) always have a solution. So the dynamical Bethe wavefunction can always be constructed if the Hilbert space of the system under consideration is small enough, for arbitrary smooth time dependence of the parameters of the model. 

In Bethe ansatz it is typical for the Bethe wavefunction to be parameterized by a number of parameters which is much smaller than size of the Hilbert space.  For example, while the Hilbert space of the XXZ Heisenberg magnet has an exponentially large dimension, its Bethe wavefunction is parametrized by a number of  parameters linearly proportional to the number of excitations, which provides a great advantage in terms of computational complexity. The equations (\ref{determ2}) and (\ref{determ1full}), however, become overdetermined if the dimensionality of Hilbert space $L>M+1$. Nevertheless, in principle, the existence of solutions for (\ref{determ2}) and (\ref{determ1full}) when it is overdetermined is not prohibited because the equations are nonlinear.  A trivial example of such a solution is adiabatic evolution when $\{\lambda^\sigma_1(t),...,\lambda^\sigma_M(t)\}$ is a solution of static Bethe equations at every moment.

\section{Example: Bose-Hubbard dimer}
\label{BHDimer}

We now illustrate the above method to obtain the time dynamics by applying it to the Bose-Hubbard dimer. This model provides both a simple and non-trivial example of a dynamically integrable model from the XXX class. For this particular case the dynamical Bethe equations can be written in a particularly simple and explicit form.


The Bose-Hubbard dimer in the two mode approximation (equivalent to a two-site Bose-Hubbard model) can be described by the following Hamiltonian \cite{milburn1997quantum,ermakov2018high}
\begin{align}
\mathcal{\hat{H}} = & \epsilon(a^\dagger a - b^\dagger b)-J(a^\dag b+ab^\dag) \nonumber \\
& + \frac{U}{2} \left(a^\dagger a^\dagger aa + b^\dagger b^\dagger b b\right)
+ V a^\dagger a  b^\dagger b , 
\label{hamb}                     
\end{align}
where $ a, b $ are bosonic operators for the two sites satisfying $[a,a^\dag]=[b,b^\dag]=1$.  The total number operator of particles $\hat{N}= a^\dag a+b^\dag b$ is a conserved quantity, $[\mathcal{\hat{H}},\hat{N}]=0$. For the Bethe ansatz formalism it is convenient rescale and offset the Hamiltonian by defining
\begin{equation}\label{hvh}
 \hat{H}=-\frac{1}{J}\left(\mathcal{\hat{H}}-\frac{U}{2}\hat{N}(\hat{N}-1)-\epsilon \hat{N}\right),
\end{equation}
which commutes with (\ref{hamb}). Defining the dimensionless detuning $\Delta=\frac{2\epsilon}{J}$ and coupling constant $c^2=\frac{U-V}{J}$, the Hamiltonian can be rewritten
\begin{equation}\label{ham}
\hat{H}=\Delta b^\dag b + a^\dag b+ab^\dag + c^2 a^\dag a b^\dag b ,
\end{equation}
which is the form we shall use. Using the dimensionless Hamiltonian (\ref{ham}) means that all energies are measured in units of $ J $ and time is measured in units of $ \hbar/J $.  

We now assume that the coupling constant $c$ is time-independent whereas the detuning $\Delta(t)$ continuously depends on time. We introduce the generalized creation operator $\mathbf{B}(\lambda)$ which depends on a complex parameter $\lambda$ \cite{bogoliubov2016time}
\begin{align}
\label{b_op_for_BH}
&\mathbf{B}(\lambda)=\lambda b^\dagger - \mathbf{X}, 
\end{align}
where
\begin{align}
\mathbf{X} = \frac{\Delta}{c} b^\dagger +ca^\dagger ab^\dagger+c^{-1}a^\dagger.
\end{align}
The pseudo-vacuum in this case is simply the zero particle Fock state
\begin{align}
\vac=|0\rangle_a\otimes|0\rangle_b. 
\end{align}

The time-dependent Bethe wavefunction can then be written following (\ref{bWF_gen}) using the above definitions. We first consider the case where only the detuning is time-dependent.  In this case the dynamical Bethe equations can be written in a compact form given by
\begin{equation}\label{dynamicalBE_BH}
i\left(\frac{\dot{\Delta}}{c}-\dot{\lambda}^\sigma_n(t)\right)=\varphi_n(\{\lambda^\sigma\})\lambda^\sigma_n(t), \; \forall n=1,...,N  .
\end{equation}
Here $\varphi_n(\{\lambda^\sigma\})$ are the so-called off-shell functions defined as
\begin{align}
\varphi_n (\{\mathbf{\lambda^\sigma}\})=& \left(\frac{\Delta}{c}-\lambda^\sigma_n\right)\prod^N_{\substack{j=1\\j\neq n}}\left(1-\frac{c}{\lambda^\sigma_n-\lambda^\sigma_j}\right)\nonumber\\
&+\frac{1}{c\lambda^\sigma_n}\prod^N_{\substack{j=1\\j\neq n}}\left(1-\frac{c}{\lambda^\sigma_j-\lambda^\sigma_n}\right).
\end{align}
When $\varphi_n(\{\mathbf{\lambda}\})=0, \; \forall n=1,...,N$, where $N$ is the number of particles, these reduce to the static Bethe equations.

The dynamical Bethe equations are a set of first order coupled ordinary differential equations. As the initial condition for (\ref{dynamicalBE_BH}) we need to pick a set of parameters $\{\lambda(0)\}=\{\lambda_1(0),...,\lambda_N(0)\}$, which parametrizes the initial state $|\Psi_N(0)\rangle$.  For example, if the initial state is an eigenstate, the set $\{\lambda(0)\}$ should satisfy the static Bethe equations. 

We numerically solve the set of equations (\ref{dynamicalBE_BH}) for a detuning with time dependence 
\begin{align}
\Delta(t)=\Delta_0+\cos(t^2),
\end{align}
which has a rather non-trivial non-linear and aperiodic dependence.  The initial condition was chosen to be the solution of static Bethe equations which corresponds to the ground state of (\ref{ham}). 
As the observable, we calculate the intersite coherence
\begin{align}
\nu(t)=\frac{|\langle a^\dagger b\rangle|}{N} ,
\label{coherence}
\end{align}
for the details of calculation of $\nu(t)$ see Appendix \ref{abaBH}. Figure \ref{DrivenDet}(a) shows our results. We find that the method perfectly reproduces time dynamics calculated by exact diagonalization, giving identical curves.  The method is computationally efficient the solution requires the evolution of $ N $ coupled equations.  In Fig. \ref{DrivenDet}(c) we show the stroboscopic maps of the solutions of dynamical Bethe equations (\ref{dynamicalBE_BH}), point of certain color corresponds to the value of the component of the solution of (\ref{dynamicalBE_BH}) $\lambda_j(t_k)$ at the moment $t_k$. Instead of solving (\ref{dynamicalBE_BH}) one may solve more general system of equations (\ref{determ2}),(\ref{determ1full}), which is applicable for arbitrary time dependece of both $\Delta$ and $c^2$, we checked that solutions of (\ref{determ2}),(\ref{determ1full}) does perfectly coincide with the solutions of (\ref{dynamicalBE_BH}) when only detuning is driven.




As a second example, we consider the case of a quench, when the parameters are changed suddenly from $c,\Delta$ to $c',\Delta'$.   When all the parameters of the model are constant the set of equations (\ref{dynamicalBE_BH}) become
\begin{equation}\label{free_dynamical_BE}
-i\frac{\dot{\lambda}_n(t)}{\lambda_n(t)}=\varphi_n(\{\lambda\}) \qquad \forall n=1,...,N .
\end{equation}
The set of equations (\ref{free_dynamical_BE}) describes the evolution of an initial state $|\Psi_N(0)\rangle$ with a static  Hamiltonian (\ref{ham}). The initial state can be parameterized by a Bethe vector with the set of parameter $\{\lambda^0\}$ satisfying (\ref{off_shell_BE}) for the initial parameters $ c,\Delta $.  
After the quench is performed, the Hamiltonian parameters change to $ c',\Delta'$, hence we need to establish the connection between the old wavefunction expressed in terms of $c,\Delta$, and the new one expressed in terms of $ c',\Delta'$.  The initial condition for (\ref{free_dynamical_BE}) thus is given by
\begin{equation}\label{WF_connect}
\prod_{j=1}^N\mathbf{B}^{(c',\Delta')} (\lambda_j(0))\vac=\prod_{j=1}^N\mathbf{B}^{(c,\Delta)} (\lambda_j^0)\vac.
\end{equation}
For the case that only the detuning is quenched $ c'= c $, the initial conditions for (\ref{free_dynamical_BE}) can be simply found to be
\begin{equation}\label{boundary cond}
\lambda_j(0)=\lambda_j^0-\frac{\Delta'-\Delta}{c}.
\end{equation}

In Fig. \ref{DrivenDet}(b) we plot an example solution of the intersite coherence (\ref{coherence}) from the dynamical Bethe equations (\ref{free_dynamical_BE}).  We again see that there is perfect agreement of the time dynamics with numerical results obtained from exact diagonalization.  In Fig. \ref{DrivenDet}(d) we plot stroboscopic maps for the solution of (\ref{free_dynamical_BE}) in the same fashion as we did for (\ref{dynamicalBE_BH}). 


\begin{figure*}
	\centering
	\includegraphics[width=\linewidth]{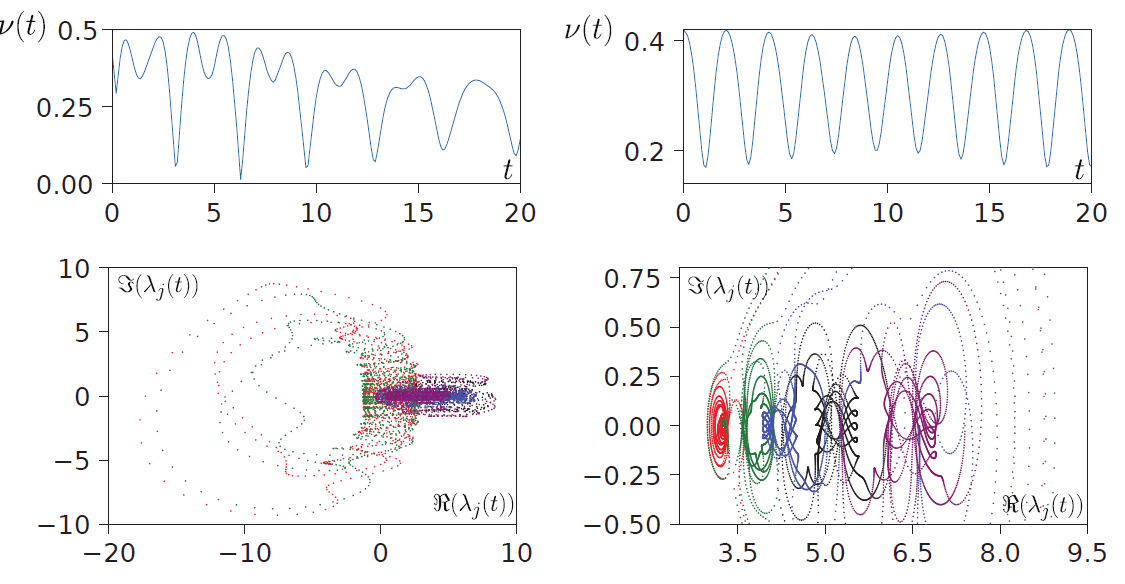}
	\caption{Time dynamics of the Bose-Hubbard dimer.  The intersite coherence (\ref{coherence}) for the case of (a) driven detuning $\Delta(t)=\Delta_0+\cos(t^2)$ using (\ref{dynamicalBE_BH}); and (b) quenched detuning from $\Delta_0=0.697$ to $\Delta'=1.697$ using (\ref{free_dynamical_BE}). Stroboscopic maps for the case of (c) driven detuning as in (a); (d) quench as in (b).  Each set of 5 points represents a solution of the dynamical Bethe equations at some certain moment of time. $N=5$ and $c=0.531$ throughout. }
	\label{DrivenDet}
\end{figure*}

\section{Outlook and conclusions}
\label{outlook}

We have described a method for evaluating the time dynamics of systems that are exactly solvable by Bethe ansatz.  
The method is based on the  dynamical Bethe wavefunction (\ref{generalB2}) which is a straightforward generalization of Bethe ansatz for dynamical case, where the Bethe parameters are time dependent and there is a time varying complex phase. The main advantage of the dynamical Bethe wavefunctions (\ref{generalB2}) is that they are mathematically manageable thanks to the well-developed Bethe ansatz results which are directly applicable. 

The set of differential dynamical Bethe equations (\ref{determ2}) and (\ref{determ1full}) can be applied to any Bethe ansatz solvable model from XXX, XXZ or Gaudin class which has a dimensionality not bigger than $N+1$, where $N$ is the number of parameters in the Bethe wavefunctions.  What would be interesting is if the dynamical Bethe wavefunction could describe the diabatic evolution of a non-trivial model with a larger Hilbert space than $ N + 1 $.  This would be an example of exact non-ergodic behavior which is a topic of great importance \cite{turner2018weak}. We have shown that our approach produces exact time dynamics with the Tavis-Cummings model, which possesses formally a Hilbert space of dimension $ 2^N $, but can be restricted by symmetry to a dimension $ N + 1 $.  Thus this alone does not demonstrate a completely non-trivial example. Currently, the in terms of computational advantage, the dynamical Bethe equations only provide an equivalent approach to alternative techniques, since both scale as $ N $.  However, we do not exclude the possibility that there are  
systems with larger Hilbert space for which overdetermined system (\ref{determ2}) and (\ref{determ1full}) might have a non-trivial solution.


\begin{acknowledgements}
The authors are grateful to O. Lychkovskiy for useful discussions. I. E. is supported by the Russian Science Foundation under the grant N$^{\rm o}$ 17-71-20158. 
\end{acknowledgements}

\appendix

\section{Algebraic Bethe ansatz}

Here we briefly sketch the main aspects of the algebraic Bethe ansatz  technique which are necessary for the understanding of the present paper. For an extensive review of the Bethe ansatz, we refer the reader to Refs. \cite{SlavnovLecturesArXiv,korepin1997quantum,essler2005one}. 

The cornerstone of any integrable model is the R-matrix which in this paper always takes the form
\begin{equation}  \label{R}
{\bf R}(\lambda ,\mu )=\left(
\begin{array}{cccc}
f(\mu ,\lambda ) & 0 & 0 & 0 \\
0 & g(\mu ,\lambda ) & 1 & 0 \\
0 & 1 & g(\mu ,\lambda ) & 0 \\
0 & 0 & 0 & f(\mu ,\lambda )
\end{array}
\right) .
\end{equation}
Here the entries $f(\mu ,\lambda )$ and $g(\mu ,\lambda )$ are specified for each model separately. In general, the R-matrix is the solution of the Yang-Baxter equation \cite{SlavnovLecturesArXiv}, and can take many different forms. The specific form of the R-matrix generates a family of integrable models.


In order to construct an integrable model we need to define the monodromy matrix 
\begin{equation}
\label{monodromy_matr}
T(\lambda)=\begin{pmatrix}
A(\lambda) & B(\lambda) \\ 
C(\lambda) & D(\lambda)
\end{pmatrix} ,
\end{equation}
which depends on the complex spectral parameter $\lambda$. Here, $A(\lambda), B(\lambda), C(\lambda)$ and $D(\lambda)$ are operators acting in the Hilbert space of the model under consideration, and their explicit representation depends on the model. The monodromy matrix should also satisfy the Yang-Baxter equation 
\begin{equation}
\label{yangbaxter}
R(\lambda,\mu)\otimes T(\lambda)\otimes T(\mu) = T(\mu)\otimes T(\lambda)\otimes R(\lambda,\mu).
\end{equation}



To construct the Hamiltonian of a particular integrable model, we define the trace of the monodromy matrix
\begin{equation}
\label{trace_mon_matr}
\tau(\lambda)=\text{Tr}T(\lambda)=A(\lambda)+D(\lambda).
\end{equation}
The Hamiltonian $\hat{\mathcal{H}}$ of the model may be expressed via the trace of the monodromy matrix $\tau(\lambda)$, or its derivative at some specified $\lambda=\lambda_0$. Usually it can be expressed as some elementary function of $\tau(\lambda_0)$, or as a residue of the $\tau(\lambda)$ at a particular point $\lambda_0$. 

The pseudovacuum state $\vac$ is a state from the Hilbert space of the model, which is annihilated by the operator $C(\lambda)\vac=0$. The conjugated operator also satisfies  $\vacc B(\lambda)=0$. Usually the pseudovacuum state is an eigenstate of the system, but, in general, it is not required. We also define two eigenvalue functions $a(\lambda)$ and $d(\lambda)$ according to
\begin{align}
\label{a_and_d}
&A(\lambda)\vac=a(\lambda)\vac \nonumber \\
&D(\lambda)\vac=d(\lambda)\vac .
\end{align}

The Bethe wavefunction is then defined as
\begin{equation}
\label{betheWF}
|\Psi(\{\lambda^\sigma_j\})\rangle=\prod^M_{j=1}B(\lambda^\sigma_j)\vac,
\end{equation}
where $\{\lambda^\sigma_j\}$ is the set of complex parameters $\{\lambda^\sigma_j\}=\{\lambda^\sigma_1,\lambda^\sigma_2,...,\lambda^\sigma_M\}$, and $M$ is the number of excitations in the system, and $\sigma$ labels the wavefunction. The wavefunction (\ref{betheWF}) is an eigenfunction of the trace of monodromy matrix $\tau(\lambda)$ 
\begin{equation}
\label{eigen_transfer}
\tau(\lambda)|\Psi(\{\lambda^\sigma_j\})\rangle=\Theta(\lambda, \{\lambda^\sigma_j\})|\Psi(\{\lambda^\sigma_j\})\rangle,
\end{equation}
if the set $\{\lambda^\sigma_j\}$ satisfies to the set of Bethe equations
\begin{equation}
\label{bethe_equations}
\frac{a(\lambda^\sigma_j)}{d(\lambda^\sigma_j)}\prod^M_{n=1\atop n\neq j}\frac{f(\lambda^\sigma_j,\lambda^\sigma_n)}{f(\lambda^\sigma_n,\lambda^\sigma_j)}=1, \qquad j=1,2,...,M.
\end{equation}
All the roots within one solution $\{\lambda^\sigma_j\}$ should be different, otherwise $|\Psi(\{\lambda^\sigma_j\})\rangle$ can not be an eigenfunction. The Bethe equations (\ref{bethe_equations}) are set of coupled nonlinear algebraic equations. It has $M$ equations and $N$ solutions, where $N$ is the size of Hilbert space.\\

For our purposes it is very important to know what the effect of the transfer matrix $\tau(\lambda)$ acting on the Bethe vector (\ref{betheWF}). For  notational simplicity we omit the index $\sigma$ henceforth, such that $\{\lambda_j\}$ denotes the set $\{\lambda_1^\sigma ,\lambda_2^\sigma,...,\lambda_M^\sigma \}$.  From Ref. \cite{SlavnovLecturesArXiv} it is known that
\begin{align}
A(\lambda) & \prod^M_{j=1}B(\lambda_j)\vac =a(\lambda)\Lambda(\lambda, \{\lambda_j\}) \prod^M_{j=1}B(\lambda_j)\vac \nonumber \\
& + \sum^M_{n=1}a(\lambda_n)\Lambda_n(\lambda, \{\lambda_j\})B(\lambda) \prod^M_{j=1 \atop j\neq n}B(\lambda_j)\vac  .
\label{a_acts_on_WF}
\end{align}
Here  we defined the functions
\begin{align}
\label{LambdaLarge}
\Lambda(\lambda, \{\lambda_j\})& =\prod^M_{j=1}f(\lambda,\lambda_j) \nonumber \\
\Lambda_n(\lambda, \{\lambda_j\})& =g(\lambda_n,\lambda)\prod^M_{j=1 \atop j\neq n}f(\lambda_n,\lambda_j) .
\end{align}
For the $D(\lambda)$ operator, we have the similar expressions
\begin{align}
D(\lambda) & \prod^M_{j=1}B(\lambda_j)\vac =d(\lambda)\bar{\Lambda}(\lambda, \{\lambda_j\}) \prod^M_{j=1}B(\lambda_j)\vac \nonumber \\
& + \sum^M_{n=1}d(\lambda_n)\bar{\Lambda}_n(\lambda, \{\lambda_j\})B(\lambda) \prod^M_{j=1 \atop j\neq n}B(\lambda_j)\vac ,
\label{d_acts_on_WF}
\end{align}
where we defined
\begin{align}
\bar{\Lambda}(\lambda, \{\lambda_j\})& =\prod^M_{j=1}f(\lambda_j,\lambda) \nonumber \\
\bar{\Lambda}_n(\lambda, \{\lambda_j\}) & =g(\lambda,\lambda_n)\prod^M_{j=1 \atop j\neq n}f(\lambda_j,\lambda_n) .
\label{LambdaLargeBar}
\end{align}
Combining these results we can find the effect of acting $\tau(\lambda)$ on the Bethe wavefunction, given by
\begin{align}
\tau(\lambda) & \prod^M_{j=1}B(\lambda_j)\vac =\Theta(\lambda, \{\lambda_j\}) \prod^M_{j=1}B(\lambda_j)\vac  \nonumber \\
& + \sum^M_{n=1}\phi_n(\lambda, \{\lambda_j\})B(\lambda) \prod^M_{j=1 \atop j\neq n}B(\lambda_j)\vac .
\label{tau_acts_on_WF}
\end{align}
Here we defined
\begin{equation}
\label{thetaL}
\Theta(\lambda, \{\lambda_j\})=a(\lambda)\Lambda(\lambda, \{\lambda_j\})+d(\lambda)\bar{\Lambda}(\lambda, \{\lambda_j\}),
\end{equation}
and the off-shell function as
\begin{equation}
\label{offshellfunc}
\phi_n(\lambda, \{\lambda_j\})=a(\lambda_n)\Lambda_n(\lambda, \{\lambda_j\})+d(\lambda_n)\bar{\Lambda}_n(\lambda, \{\lambda_j\}).
\end{equation}
If we now demand that the off-shell function (\ref{offshellfunc}) is zero, it is evident that the wavefunction (\ref{betheWF}) is an eigenfunction for $\tau(\lambda)$. The roots of off-shell functions (\ref{offshellfunc}) coincide with the roots of Bethe equations (\ref{bethe_equations}), but we should distinguish between these since later we will encounter cases where the off-shell function is not zero.

Finally, we mention several important properties of Bethe wavefunctions. The dual Bethe wavefunctions are defined as
\begin{equation}
\label{dual_betheWF}
\langle\Psi(\{\lambda^\sigma_j\})|=\vacc\prod^M_{j=1}C(\lambda^\sigma_j).
\end{equation}
In general, despite the notation, the wavefunction (\ref{dual_betheWF}) does not coincide with the hermitian conjugate of the function (\ref{betheWF}), i.e.  $\langle\Psi(\{\lambda^\sigma_j\})|\neq |\Psi(\{\lambda^\sigma_j\})\rangle^\dagger$. Dual vectors like this must be introduced in order to evaluate scalar products and averages of observables. Generally, in the literature devoted to Bethe Ansatz, the left bracket $\langle\Psi|$ implies the dual vector (\ref{dual_betheWF}). 

For most of the integrable models it has been proven that Bethe vectors form a complete set \cite{korepin1997quantum,SlavnovLecturesArXiv}
\begin{equation}
\label{complete_set_BWF}
\sum^N_{\sigma=1}|\Psi(\{\lambda^\sigma_j\})\rangle\langle\Psi(\{\lambda^\sigma_j\})|\propto \hat{I} ,
\end{equation}
where $\hat{I}$ is the identity operator, and $N$ is the size of the Hilbert space. In general Bethe wavefunctions are not normalized. 

One of the most important properties of Bethe wavefunctions is that for many models it is possible to evaluate the scalar product of Bethe wavefunctions and averages of the operators by applying Slavnov's formula \cite{slavnov1989calculation}.  This allows one to express scalar product as a determinant. We do not reproduce the general form of the Slavnov's formula here because of its complexity, and it not very useful to consider it without specifying the model. Application of Slavnov's formula to the models considered in this paper have been studied in Refs. \cite{bogoliubov2016time,bogoliubov2017time}.

\section{Dynamical Bethe equations for the Bose-Hubbard dimer}
\label{abaBH}

Here we give more details of the derivation of the dynamical Bethe equations for the detuning driven Bose-Hubbard dimer.  A more detailed description regarding the Bethe ansatz solution of this model can be found in Ref. \cite{bogoliubov2016time}, we use the same notations as this paper. 

The Hamiltonian of the Bose-Hubbard dimer is
\begin{equation}\label{hamsupp}
\hat{H}=\Delta b^\dag b + a^\dag b+ab^\dag + c^2 a^\dag a b^\dag b .
\end{equation}
The diagonal elements of the monodromy matrix are in this case
\begin{align}
\label{a_and_d_ham}
A(\lambda) =& \lambda^2-\lambda\left(ca^\dagger a+cb^\dagger b+\frac{\Delta}{c}\right) \nonumber \\
& +\Delta b^\dagger b+a^\dagger b+c^2a^\dagger ab^\dagger b, \\
D(\lambda)  = & ab^\dagger+c^{-2}.
\end{align}
The Hamiltonian (\ref{hamsupp}) can then be expressed via trace of the monodromy matrix (\ref{trace_mon_matr}) according to
\begin{equation}
\label{ham_via_tr}
\hat{H}=\tau(0)-c^{-2}.
\end{equation}
According to the definitions (\ref{a_and_d}), the eigenvalue functions are then
\begin{align}
\label{a_and_d__smham}
a(\lambda) &=\lambda\left(\lambda-\frac{\Delta}{c}\right), \\
d(\lambda)& = c^{-2}.
\end{align}
The elements of the R-matrix are defined as
\begin{align}
\label{f_and_g_ham}
&f(\mu,\lambda)=1-\frac{c}{\mu-\lambda} \\
&g(\mu,\lambda) = -\frac{c}{\mu-\lambda}.
\end{align}

We now wish to look for Bethe eigenfunctions of the form
\begin{equation}\label{bWF_BH}
|\Psi^\sigma_N\rangle=\prod^N_{j=1}\mathbf{B}(\lambda^\sigma_j)\vac,
\end{equation}
where the pseudo-vacuum state is $\vac=|0\rangle_a\otimes|0\rangle_b$.   $\sigma$ is the index which labels the energy levels of the system, for the sake of notational simplicity we omit this below. The eigenvector depends on $N$ complex parameters $\{\lambda\}=\{\lambda_1,\lambda_2,...,\lambda_N\}$. By applying the Bethe ansatz machinery we can evaluate
\begin{align}
\label{h_on_bWF}
\hat{H}|\Psi_N\rangle= & E_N(\{\lambda\})\prod^N_{j=1}\mathbf{B}(\lambda_j)\vac \nonumber \\
& -\sum_{n=1}^N\varphi_n(\{\mathbf{\lambda}\})\mathbf{X}\prod_{\substack{j=1\\j\neq n}}^N\mathbf{B}(\lambda_j)\vac,
\end{align}
where we have defined
\begin{align}
\label{b_op_for_BH}
E_N(\{\lambda\}) =&-c^{-2}+c^{-2}\prod^N_{j=1}\left(1-\frac{c}{\lambda_j}\right), \\
\varphi_n(\{\mathbf{\lambda}\})=&  \left(\frac{\Delta}{c}-\lambda_n\right)\prod^N_{\substack{j=1\\j\neq n}}\left(1-\frac{c}{\lambda_n-\lambda_j}\right) \nonumber \\
& +\frac{1}{c\lambda_n}\prod^N_{\substack{j=1\\j\neq n}}\left(1-\frac{c}{\lambda_j-\lambda_n}\right).
\end{align}
Here $E_N(\{\lambda\})$ is the energy and $\varphi_n(\{\mathbf{\lambda}\})$ is the off-shell function. From (\ref{h_on_bWF}) we can see that when set $\{\lambda\}$ satisfies 
\begin{equation}\label{off_shell_BE}
\varphi_n(\{\mathbf{\lambda}\})=0, \qquad \forall n=1,...,N ,
\end{equation}
the wavefunction (\ref{bWF_BH}) becomes an eigenfunction of the Hamiltonian (\ref{hamsupp}). The set of equations (\ref{off_shell_BE}) are known as the Bethe equations. 

We now look for a time-dependent wavefunction of the form
\begin{equation}\label{timed_bwf_BH}
|\Psi_N(t)\rangle=e^{ip(t)}\prod^N_{j=1}\mathbf{B}(\lambda_j(t))\vac .
\end{equation}
If only the detuning $ \Delta $ is time-dependent, it is easy to see that $[\frac{d}{dt}\mathbf{B},\mathbf{B}]=0$, and the derivative of (\ref{timed_bwf_BH}) can be taken easily.  Substituting (\ref{timed_bwf_BH}) into the time-dependent Schrodinger equation we obtain 
\begin{align}\label{psitimeoff}
[p'(t)+&E_N(\{\mathbf{\lambda}\})]\prod_{j=1}^N\mathbf{B}(\lambda_j)\vac=\nonumber\\
&\sum_{n=1}^N\left(i\left(\dot{\lambda}_n-\frac{\dot{\Delta}}{c}\right)b^\dagger+\varphi_n(\{\mathbf{\lambda}\})\mathbf{X}\right)\prod_{\substack{j=1\\j\neq n}}^N\mathbf{B}(\lambda_j)\vac .
\end{align}
If we demand now that
\begin{equation}\label{dynamicalBE_BH_app}
i\left(\frac{\dot{\Delta}}{c}-\dot{\lambda}_n(t)\right)=\varphi_n(\{\lambda\})\lambda_n(t), \qquad \forall n=1,...,N \;,
\end{equation}
the wavefunction (\ref{timed_bwf_BH}) will satisfy the time-dependent Schrodinger equation. We call the set of conditions (\ref{dynamicalBE_BH}) the dynamical Bethe equations. The dynamical Bethe equations are set of first order coupled ordinary differential equations. For the initial condition of (\ref{dynamicalBE_BH}) we need to pick a set $\{\lambda(0)\}=\{\lambda_1(0),...,\lambda_N(0)\}$, which parametrizes the initial state $|\Psi_N(0)\rangle$.  For example if the initial state is an eigenstate, the set $\{\lambda(0)\}$ should satisfy the static Bethe equations (\ref{off_shell_BE}). 
The phase factor $p(t)$ is given by 
\begin{equation}\label{phasefactor_BH}
p(t)=\int_0^tdt'\left(-E_N(\{\mathbf{\lambda}\})+\sum_{n=1}^N\frac{i}{\lambda_n}\left(\dot{\lambda}_n-\frac{\dot{\Delta}}{c}\right)\right).
\end{equation} 

To evaluate observables one may use the determinant representation as a general approach \cite{slavnov1989calculation,bogoliubov2016time}. More convenient approach is to use the expansion of Bethe vectors (\ref{bWF_BH}) over the Fock space, which was developed in \cite{ermakov2018high}:

\begin{align}\label{wavefExp}
\nonumber|\Psi_N(\{\lambda\})\rangle&=\sum _{m=0}^N \sum _{l=0}^{N-m} \sum _{k=0}^l(-1)^m \sqrt{k!} \sqrt{(N-k)!} D(l,k)\\
\nonumber&\binom{N-m}{l} \Gamma _{lmk}|k\rangle_a\otimes|N-k\rangle_a,\\
\nonumber\langle\Psi_N(\{\lambda\})|&=\sum _{m=0}^N \sum _{k=0}^{N-m}(-1)^m\langle N-k|_a\otimes\langle k|_b \sqrt{k!}\\
&\sqrt{(N-k)!} c^{-2 k-m+N} D(N-m,k)e_m,
\end{align}
where the coefficient $\Gamma _{lmk}$ defined as
\begin{equation}\label{gammacoef}
\Gamma _{lmk}=\Delta^{N-m-l}c^{-N+m+2l-2k}e_m,
\end{equation}
and $D(M,k)$ are coefficients defined by the following recurrence relation
\begin{equation}\label{recurrent}
D(M,k)=kD(M-1,k)+D(M-1,k-1)
\end{equation}
with the conditions: $D(1,1)=1$ and $D(M,k)=0$ if $k>M$. This coefficient possess the obvious property: $D(M,1)=D(n,n)=1$. The general expression for $D(M,k)$ is given by
\begin{multline}\label{dcoeff}
D(M,k)=\sum\limits^{M-k}_{n_1=0}\sum\limits^{M-k-n_1}_{n_2=0}\sum\limits^{M-k-n_1-n_2}_{n_3=0}...\\\sum\limits^{M-k-n_1-...-n_{k-1}}_{n_{k-1}=0}k^{n_1}(k-1)^{n_2}\;...\;2^{n_{k-1}}.
\end{multline}

%
%
%
\bibliographystyle{apsrev}
\bibliography{ref}

\end{document}